\documentclass{mem}
\usepackage{natbib}\usepackage{txfonts}\usepackage{balance}
\usepackage{graphicx}
\usepackage[a4paper]{hyperref}
\idline{75}{282}
\begin{document}
\def\teff{$T\rm_{eff }$}
\def\kms{$\mathrm {km s}^{-1}$}

\title{MCAO for the European Solar Telescope: first results}

   \subtitle{}

   \author{M. Stangalini, R. Piazzesi, D. Del Moro, F. Berrilli, A. Egidi}
   \institute{Physics Department, University of Rome Tor Vergata\\
   \email{marco.stangalini@roma2.infn.it}}
   \offprints{M. Stangalini}

\authorrunning{M. Stangalini \emph{et al.}}
\titlerunning{EST-MCAO}

\abstract{
We analize the efficiency of wavefront reconstruction in the MultiConjugate Adaptive Optics system for the European Solar Telescope (EST).
We present preliminary results derived from numerical simulations.
We study a 4 meter class telescope with multiple deformable mirrors conjugated at variable heights. 
Along with common issues, difficulties peculiar to the solar case have to be considered, such as the low contrast and extended nature of the natural guide features.
Our findings identify basic requirements for the EST Adaptive Optics system and show some of its capabilities.

}
\maketitle{}

\section{Introduction}
The European Solar Telescope (EST), currently in its design study phase, will be a 4-meter class solar telescope. EST will be optimized for the study of the magnetic coupling between the deep photosphere and the upper chromosphere. The main observational focus will be multi-wavelenght (spectro-)polarimetry at photospheric and chromospheric layers.\\
Plasma phenomena on the solar scene seem to happen on a small scale ($<100km$) and to exhibit high temporal evolution (seconds, minutes), thus high spatial and temporal resolutions are needed.\\
To fit these needs, EST will be equipped with a high technology Multi Conjugate Adaptive Optics (MCAO) system able to deliver high resolution images of the Sun with a corrected Field of View (FoV) of $1$ arcmin \citep{SRD}.\\
It this paper we will report the first results of our MCAO simulations, using a layer oriented approach \citep{Ragazzoni}.\\
In particular, we have studied how the corrected FoV depends on the configuration of the system tracking points on the solar surface and furthermore how it depends on the modal base dimension employed in the wavefront reconstruction.\\
Finally we have tested through simulations a time delay reduction method based on wavefront forecasting. Its use seems to increase the performance of the MCAO system and thus the quality of the correction.

   \begin{figure*}[]
   \centering
   \includegraphics[width=14cm]{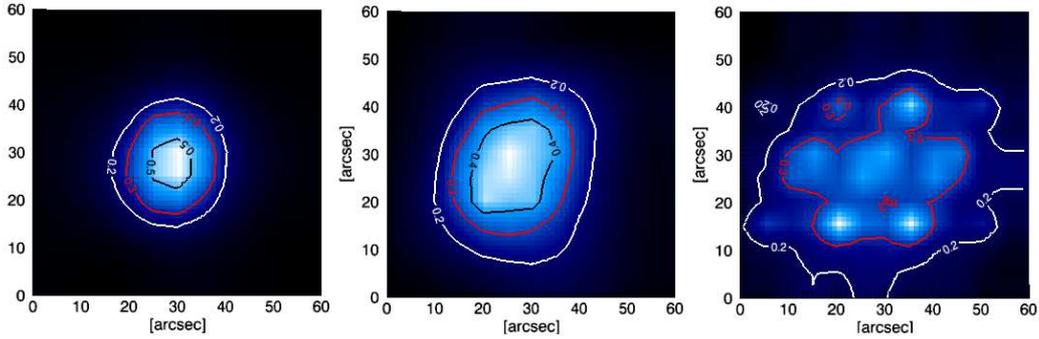}
      \caption{SR maps for three different asterism radius: $7$ (left), $10$ (center) and $15$ (right) $arcsec$. The best performance is achieved for the $10$ $arcsec$ radius for which the Strehl Ratio reaches $0.5$ over a fairly large FoV ($20$ $arcsec$) and a value of $0.3$ over $50$ $arcsec$. In the right panel it is possible to see how further increasing the asterism radius produces an unsatisfactory correction in terms of FoV coverage and SR peak. Map FoV is equal to $1~arcmin~x~1~arcmin$}
   \end{figure*}

\section{MCAO simulations}
In this work we have analized the MCAO system behaviour using the LOST simulation tool \citep{Arcidiacono} along with some optimization to include needs typical of the solar case and wavefront prediction capabilities.\\
As mentioned above, we have used simulations to study the corrected FoV and the quality of the correction itself.\\
To do this, we have simulated a MCAO system with the following characteristics:
\begin{itemize}
 \item $2$ Deformable Mirrors (DM)
 \item Conjugation altitudes: $0~Km$, $10~Km$
 \item Time step: $0.5~ms$
 \item Loop delay: $1~ms$
 \item Working wavelength: $555~nm$
 \item $7$ Shack-Hartmann (SH) wavefront sensors (WFS) with $30x30$ subpupils sampling
 \item $2$ Kolmogorov turbulence layers (FFT): $0~km$, $10~km$
 \item $D/r_0$ : $26$, $8$
\end{itemize}
where $D$ is the telescope diameter and $r_0$ the Fried parameter.\\
In these simulations we have used a hexagonal tracking point geometry (asterism).\\
We have studied, at first, sky coverage dependence on asterism radius, using as a quality descriptor the Strehl Ratio parameter (SR):

\begin{equation}
 SR=\frac{PSF_{peak}}{PSF_{diffraction~limited~peak}}
\end{equation}

In fig. $1$ we present SR maps for three different choices of the asterism radius: $7$, $10$ and $15$ $arcsec$.

As is clear, the best correction is achieved when using a $10$ $arcsec$ asterism radius, where the largest sky coverage is reached with a $20$ $arcsec$ FoV and $SR=0.3$.\\ 
A $SR=0.3$ is generally considered, in terms of spatial resolution, as a fairly good condition where the PSF width is very close to the diffraction limited one.\\
Every attempt to enlarge the corrected FoV by increasing the asterism radius produces an unsatisfactory condition with a SR that is not uniform over the field and with a decrease of the SR peak itself (see right panel of fig.1). Thus there exists a maximum useful asterism radius, above which the correction becomes unstable.\\
Fig. 2 and 3 show, in fact, the maximum SR reached in every simulation iteration (time step = $0.5~ms$) over the whole FoV for a $10$ and $15$ $arcsec$ asterism radius respectively. While in the first case the correction is stable, in the second case it is unstable: after $200$ loops the SR peak decreases rapidly. 
  
 \begin{figure}[]
   \centering
   \includegraphics[width=6cm]{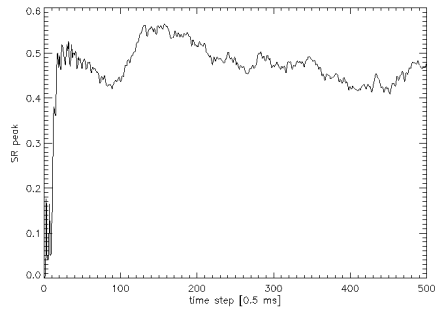}
      \caption{SR peak as a function of time for a $10~arcsec$ asterism radius. Every iteration is equal to $0.5~ms$. The correction is stable showing a fairly constant SR peak value.}
   \end{figure}

The second case also shows a lower SR mean value with respect to the case in which the asterism radius is $10~arcsec$.

 \begin{figure}[]
   \centering
   \includegraphics[width=6cm]{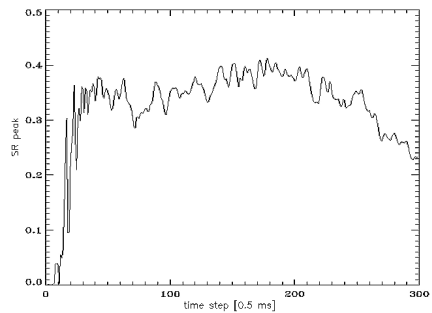}
      \caption{SR peak as a function of time for a $15~arcsec$ asterism radius. Every iteration is equal to $0.5~ms$. The correction is poorly stable and after $200$ loops the SR dramatically decreases.}
   \end{figure}

\section{Time delay error reduction}
One of the most important problems occurring in the MCAO correction is the time delay error.\\
Every MCAO system takes some milliseconds to analize the incoming wavefront and compute the command matrix to move the mirror. This means that the correction doesn't exactly match the incoming wavefront, leaving some residuals which can decrease the SR and the sky coverage.\\
One attempt to solve this problem is to use short-term wavefront forecasting.
We have implemented a prediction tool algorithm based on linear extrapolation inside LOST. This algorithm is able to achieve a prediction of the modal coefficients starting from their respective known time series.\\
In this way we have been able to simulate the behaviour of the prediction tool during the loop closure.\\
The algorithm makes use of the last two modal coeffiecient values to estimate the short-term predicted one.\\
We have investigated the behaviour of the SR with and without the prediction tool with the same simulations parameters.
The results are shown in fig. 4.\\

 \begin{figure}[]
   \centering
   \includegraphics[width=6cm]{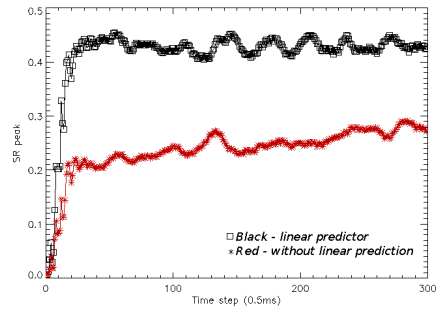}
      \caption{SR plot as a function of time step. Comparison of SR without (red asterisks) and with (black boxes) linear prediction. When the linear predictor is operative the SR reaches a value of about $0.4$ which is double the value of the case without prediction.}
   \end{figure}

As is clear, the SR is enhanced by the prediction, reaching a mean value of $0.4$. In the same plot we show, for comparison, the performance of the correction without the linear prediction of modal coefficients: in this case the SR reaches $0.2$ when the loop is closed.

\section{Modal base optimization using information theory}
Another way to minimize the time delay effects is by otpimizing the modal base. If it is possible to use fewer modes in the correction then the calculation will be speeded up, leading to a reduction in the time delay.\\
In this work we have optimized the standard Zernike modal base using information theory, working on real data obtained at the German VTT telescope \citep{vdl}.\\
We have reordered the modal base by mutual information \citep{haykin}, forcing those modes which bring more information about the incoming wavefront to appear in the first positions. We have then estimated the fitting error comparing it to that obtained in the case of standard ordering.

 \begin{figure}[]
   \centering
   \includegraphics[width=7cm]{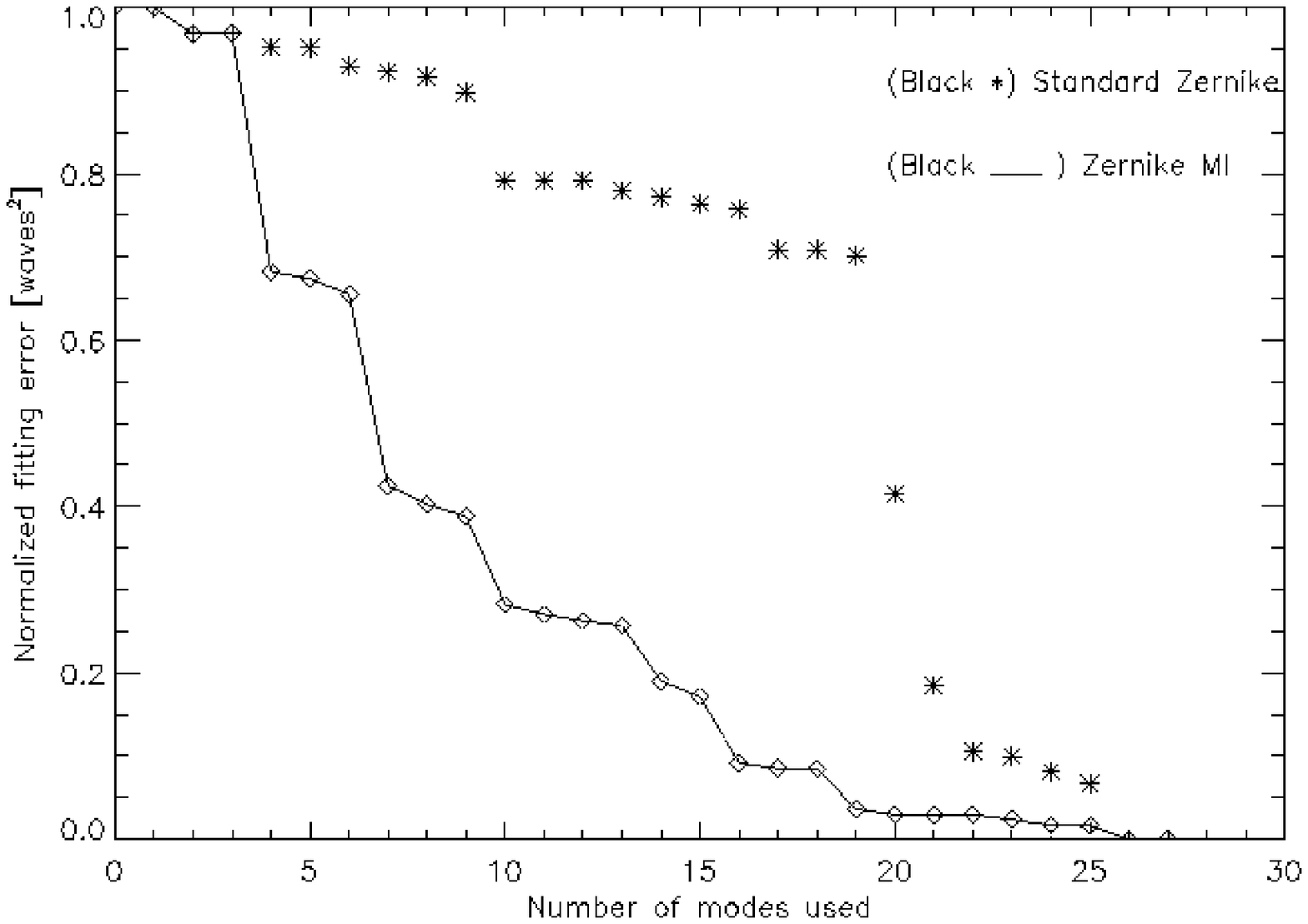}
      \caption{Fitting error as a function of the number of modes used in the reconstruction. Mutual information Zernike reordering (connected points) leads to a faster decreasing trend with respect to standard modal ordering (asterisk). For a given fitting error we can use far fewer modes in the case of mutual information ordering.}
   \end{figure}

The results of this analysis are shown in fig. 5. Mutual information ordering leads to a faster decrease of the fitting error when compared to the standard Zernike ordering. In this way, for a given fitting error, we can choose to use fewer modes, thus speeding up the calculation.

\end{document}